\input{psfig.sty}
\documentclass[oldversion]{aa}
\usepackage{maple2e}

\newcommand{\be}{\begin{equation}}
\newcommand{\ee}{\end{equation}}
\newcommand{\bea}{\begin{eqnarray}}
\newcommand{\eea}{\end{eqnarray}}
\def\kms {\ km s$^{-1}$}
\def\msol{\ifmmode {\>M_\odot}\else {$M_\odot$}\fi}
\def\cmsq{\ifmmode {\>{\rm\ cm}^2}\else {cm$^2$}\fi}
\def\psqcm{\ifmmode {\>{\rm cm}^{-2}}\else {cm$^{-2}$}\fi}
\def\psqpc{\ifmmode {\>{\rm pc}^{-2}}\else {pc$^{-2}$}\fi}
\def\pcsq{\ifmmode {\>{\rm\ pc}^2}\else {pc$^2$}\fi}
\def\Tkev{\ifmmode{T_{\rm kev}}\else {$T_{\rm keV}$}\fi}
\def\hubunits{\ifmmode {\>{\rm km\ s^{-1}\ Mpc^{-1}}}\else {km
s$^{-1}$ Mpc$^{-1}$}\fi}
\def\gta{\;\lower 0.5ex\hbox{$\buildrel > \over \sim\ $}}
\def\lta{\;\lower 0.5ex\hbox{$\buildrel < \over \sim\ $}}
\def\phiIV{\ifmmode{\varphi_4}\else {$\varphi_4$}\fi}
\def\phiI{\ifmmode{\varphi_i}\else {$\varphi_i$}\fi}
\def\be{\begin{equation}}
\def\ee{\end{equation}}
\def\bea{\begin{eqnarray}}
\def\eea{\end{eqnarray}}
\def\beas{\begin{eqnarray*}}
\def\eeas{\end{eqnarray*}}
\def\gtrapprox{\;\lower 0.5ex\hbox{$\buildrel >\over \sim\ $}}
\def\lessapprox{\;\lower 0.5ex\hbox{$\buildrel < \over \sim\ $}}

\def\Em    {${\cal E}_m$}

\def\tauLL {\ifmmode{\tau_{\scriptscriptstyle LL}}\else 
           {$\tau_{\scriptscriptstyle LL}$}\fi}

\def\Em{\ifmmode{{\rm E}_m}\else {{\rm E}$_m$}\fi}
\def\NH{\ifmmode{{\rm N}_{\scriptscriptstyle\rm H}}\else {{\rm N}$_{\scriptscriptstyle\rm H}$}\fi}

\def\eg    {{\it e.g.}}

\def\etal  {\ et al.}
\def\kms{\ifmmode {\>{\rm\ km\ s}^{-1}}\else {\ km s$^{-1}$}\fi}
\def\Em{\ifmmode{{\cal E}_m}\else {{\cal E}$_m$}\fi}
\def\Dm{\ifmmode{{\cal D}_m}\else {{\cal D}$_m$}\fi}
\def\fesc{\ifmmode{\hat{f}_{\rm esc}}\else {$\hat{f}_{\rm esc}$}\fi}
\def\fescs{\ifmmode{f_{\rm esc}}\else {$f_{\rm esc}$}\fi}
\def\rsolar{\ifmmode{r_\odot}\else {$r_\odot$}\fi}
\def\emunit{\ifmmode{{\rm cm}^{-6}{\rm\ pc}}\else {
cm$^{-6}$ pc}\fi}
\def\intensity{\ifmmode{{\rm erg\ cm}^{-2}{\rm\ s}^{-1}
      {\rm\ Hz}^{-1}{\rm\ sr}^{-1}}
      \else {erg cm$^{-2}$ s$^{-1}$ Hz$^{-1}$ sr$^{-1}$}\fi}
\def\flux{\ifmmode{{\rm erg\ cm}^{-2}{\rm\ s}^{-1}}\else {erg
cm$^{-2}$ s$^{-1}$}\fi}
\def\fluxdensity{\ifmmode{{\rm erg\ cm^{-2}\ s^{-1}\ Hz^{-1}}}\else {erg
cm$^{-2}$ s$^{-1}$ Hz$^{-1}$}\fi}
\def\phoflux{\ifmmode{{\rm phot\ cm}^{-2}{\rm\ s}^{-1}}\else {phot
cm$^{-2}$ s$^{-1}$}\fi}
\def\phorate{\ifmmode{{\rm phot\ s}^{-1}}\else {phot s$^{-1}$}\fi}
\def\apj{{\it Ap.J.~}}
\def\apjs{{\it Ap.J.Suppl.~}}
\def\apss{{\it Astrophys.Sp.Science~}}
\def\aj{{\it Astron.J.~}}

\def\etal{{\it et al.~\/}}

\def\eg{{\it e.g.}}
\def\ltsima{$\; \buildrel < \over \sim \;$}
\def\simlt{\lower.5ex\hbox{\ltsima}}
\def\gtsima{$\; \buildrel > \over \sim \;$}
\def\simgt{\lower.5ex\hbox{\gtsima}}
\def\fesc{{$\langle f_{\rm esc}\rangle$}\xspace}
\def\h2{H$_2$\xspace}

\begin{document}

%\title{The Cusp/Core problem and the Secondary Infall Model} 
%\title{Improved constraints on the Cosmological parameters from X-ray luminous clusters}
\title{Improvements in the X-ray luminosity function 
%Improved 
and constraints on the Cosmological parameters from X-ray luminous clusters}
%\title{Haloes' shape on galactic and cluster scales} 

\author{A. Del Popolo\inst{1,2}, V. Costa\inst{3}, G. Lanzafame\inst{4}
}
\titlerunning{Cosmological Constraints}
\authorrunning{A. Del Popolo}
\date{}
\offprints{A. Del Popolo, E-mail:antonino.delpopolo@unibg.it}
\institute{
$^1$ Dipartimento di Fisica e Astronomia, Universit\'a di Catania, Viale Andrea Doria 6, 95125 Catania, Italy\\
$^2$ Argelander-Institut f\"ur Astronomie, Auf dem H\"ugel 71, D-53121 Bonn, Germany\\
%$3$ Istanbul Technical University, Ayazaga Campus,  Faculty of Science and Letters,  34469 Maslak/ISTANBUL, Turkey\\
$^3$ Dipartimento di Metodologie Chimiche e Fisiche per l'Ingegneria, Univesita di Catania, Viale
A. Doria 6, I-95125, Catania, Italy\\
$^4$ Osservatorio Astrofisico di Catania, Istituto Nazionale di Astrofisica, Via S. Sofia 78, I-95123,
Catania, Italy\\
}
\abstract{
We show how to improve constraints on  
%on the mean matter density, 
$\Omega_m$, 
%the normalization of the density fluctuation power spectrum, 
$\sigma_8$, and the dark-energy equation-of-state parameter, $w$, 
obtained by Mantz et al. (2008) from measurements of the X-ray luminosity function 
of 
%the largest known 
galaxy clusters, namely MACS, the local BCS and the REFLEX galaxy cluster samples
with luminosities $L> 3 \times 10^{44}$ erg/s in the 0.1--2.4 keV band. 
%%%%%%%%%%%The main goal of the paper is to show that the XLF alone can give enough tight constraints without necessarily recurring to combination %%%%%%%%%%%%with other data sets.
%at redshifts $z<0.7$, as compiled in the Massive Cluster Survey (MACS) and the local BCS 
%and REFLEX galaxy cluster samples.
To this aim, we use Tinker et al. (2008) mass function instead of Jenkins et al. (2001) and the M-L relationship
obtained from Del Popolo (2002) and Del Popolo et al. (2005). 
Using the same methods and priors of Mantz et al. (2008), we find, for a $\Lambda$CDM universe, $\Omega_m=0.28^{+0.05}_{-0.04}$ and $\sigma_8=0.78^{+0.04}_{-0.05}$ while the result of Mantz et al. (2008) gives less tight constraints $\Omega_m=0.28^{+0.11}_{-0.07}$ and $\sigma_8=0.78^{+0.11}_{-0.13}$. In the case of a $w$CDM model, we find 
$\Omega_m=0.27^{+0.07}_{-0.06}$, $\sigma_8=0.81^{+0.05}_{-0.06}$ and $w=-1.3^{+0.3}_{-0.4}$,
while in Mantz et al. (2008) they are again less tight $\Omega_m=0.24^{+0.15}_{-0.07}$, $\sigma_8=0.85^{+0.13}_{-0.20}$ and $w=-1.4^{+0.4}_{-0.7}$.
Combining the XLF analysis with the $f_{gas}$+CMB+SNIa data set results in the constraint 
%$\Omega_m=0.269 \pm 0.015$, $\sigma_8=0.81^{\pm 0.023}$ and $w=-1.02^{\pm 0.05}$ 
$\Omega_m=0.269 \pm 0.012$, $\sigma_8=0.81 \pm 0.021$ and $w=-1.02 \pm 0.04$, 
to be compared with Mantz et al. (2008), $\Omega_m=0.269 \pm 0.016$, $\sigma_8=0.82 \pm 0.03$ and $w=-1.02 \pm 0.06$.
The tightness of the last constraints obtained by Mantz et al. (2008), are fundamentally due to the tightness 
of the $f_{gas}$+CMB+SNIa constraints and not to their XLF analysis. 
Our findings, consistent with $w=-1$, lend additional support to the cosmological-constant model. 
}
%\keywords{cosmology: theory - large scale structure of universe - galaxies:
%formation}

%\end{document}
\maketitle

\section{Introduction}

Cluster of galaxies are the
largest gravitationally-collapsed structures in the Universe. Even at the present epoch
they are relatively rare, with only a few percent of galaxies being in clusters.
In the hierarchical collapse scenario for structure formation in the universe, 
the number density of collapsed objects as a function of mass and cosmic time is a sensitive probe of cosmology. 
The galaxy clusters that occupy the
high-mass tail of this population provide a powerful and
relatively clean tool for cosmology, since their growth is
predominantly determined by linear gravitational processes.

%VIKHLININ
Starting in the 1990's, analysis of massive clusters 
%they 
have consistently indicated low values of 
$\Omega_m$ (both from the baryonic fraction arguments (White et al. 1993) and  measurements of the evolution 
in the cluster number density (Eke et al. 1998; Borgani et al. 2001) and low values of
$\sigma_8$\footnote{$\sigma_8$ is the amplitude of the mass density fluctuation power spectrum over spheres of radius $8 h^{-1} {\rm Mpc}$, and $M_8$ 
is the mean mass within these spheres} (Henry \& Arnaud 1991; Reiprich \& B\"oringer 2002; Schuecker et al. 2003) --a result since then confirmed by cosmic microwave background (CMB) studies, cosmic
shear, and other experiments (Spergel et al. 2007; Komatsu et al. 2008; Dunkley et al. 2008; Benjamin et al. 2007; Fu et al. 2008). 
For precision's sake, cluster surveys in the local universe are particularly useful
for constraining a combination of the matter density parameter 
$\Omega_m$ and the normalization of the power spectrum of density fluctuations. Following the evolution of the cluster
space density over a large redshift baseline, one can break the degeneracy between $\sigma_8$ and $\Omega_m$ (Rosati et al. 2002).
Recently, X-ray studies (Vikhlinin et al. 2009b) of the evolution of the cluster
mass function at $z =$ 0-0.8 have convincingly demonstrated that the growth of cosmic
structure has slowed down at $z < 1$ due to the effects of dark energy, and these measurements
have been used to improve the determination of the equation of state parameter.
%{\bf
Although the quoted cosmological test is very powerful, there are two main problems in practical applications: 
first, theoretical predictions provide the number density of clusters of a given mass, while the mass itself is never the directly observed quantity. Second, a cluster sample is needed that spans a large-$z$ baseline and is based on model-independent selection criteria.\footnote{This is so that the search volume and the number density associated with each cluster are uniquely identified.}
%}

Determining the evolution of the space density of clusters requires counting the
number of clusters of a given mass per unit volume at different redshifts. Therefore, three essential tools are required for its application as a cosmological test: (a) an efficient method to find clusters over a wide redshift range, (b) an observable estimator of the cluster mass, and (c) a method to compute the selection function or equivalently the survey volume within which clusters are found.
Observations of clusters in the X-ray band provide an efficient and physically
motivated method of identification, which fulfills the three requirements above.
The X-ray luminosity,  provides a very efficient method for identifying clusters down to a given X-ray flux limit and hence within  a known survey volume for each luminosity $L_x$,  which uniquely specifies the cluster selection, is also a good probe of the depth of the cluster gravitational potential. For these reasons most of the cosmological studies based on clusters have used X--ray–-selected samples.

According to the three points quoted above,
the recipe for constraining cosmological parameters by means of clusters is composed of three ingredients:
1) The predicted mass function of clusters, $n(M,z)$, as a function of cosmological parameters ($\sigma_8$, $\Omega_m$, $w$, etc.). 
%%%%%%(SPECICARE COSA SONO). 
%The mass function can be calculated analytically (METTERLI IN BIBLIOGRAFIA Press \& Schechter 1974; Sheth \& Tormen ??; Del Popolo  ?) or fitting %the results of numerical simulations (METTERLI IN BIBLIOGRAFIA Jenkins et al. 2001; Reed et al. 2003, Yahagi et al. 2004; Tinker et al. 2008). \\
2) Sky surveys with well understood selection functions to find clusters, as well as a
relation linking cluster mass with an observable. A successful solution to the former requirement has been to identify
clusters by the X-ray emission produced by hot intracluster gas, notably using data from 
%the 
ROSAT\footnote{The ROSAT Brightest Cluster Sample (BCS; Ebeling et al. 1998, 2000) and ROSAT-ESO Flux Limited X-ray sample (REFLEX; B\"ohringer et al. 2004) together cover approximately two-thirds of the sky out to redshift $z \simeq 0.3$ and contain more than 750 clusters. The Massive Cluster Survey (MACS; Ebeling et al. 2001, 2007)
%–- which at this writing contains 126 clusters and covers 55 per cent of the sky – 
extends these data to $z \simeq 0.7$.
The ROSAT 160 sq. degree survey, described for the first time by Vikhlinin et al. (1998)
%(Vikhlinin et al. 2009), 
is a serendipitous cluster catalogue containing 201 groups/clusters, while the ROSAT  
400 sq. degree survey is based on 1610 high Galactic latitude ROSAT PSPC pointings (Burenin et al. 2007)
and includes 266 optically confirmed galaxy clusters, groups and individual elliptical galaxies.
}. 
%All-Sky Survey (RASS; Tr\"umper 1993). 
%
%A large, wide-area, clean, complete cluster survey, with a well defined selection function. 
%Current leading work based on ROSAT X-ray surveys. 
%
%Future important work based on next-generation X-ray surveys (eROSITA/WFXT)   
%as well as new SZ (e.g. SPT, Planck, ACT and optical catalogues). 
3) A tight, well-determined scaling relation between survey observable (e.g. $L_x$) and mass, with minimal intrinsic scatter. \\

Early attempts to use evolution of the cluster mass function as a cosmological probe were limited by small sample sizes and either poor proxies for the cluster mass (e.g., the total X-ray flux) or inaccurate measurements (e.g. temperatures with large uncertainties)

Until some years ago the obtained results for $\Omega_{\rm m}$ were several times in disagreement. Study by different authors (Bahcall, Fan \& Cen (1997), Bahcall \& Fan (1998), Sadat, Blanchard \& Oukbir (1998), Blanchard, Bartlett \& Sadat (1998), Blanchard \& Bartlett (1998), Eke et al. (1998), Viana \& Liddle (1999), Reichart et al. (1999), Donahue \& Voit 1999, Borgani et al. 2001) found values for $\Omega_m$
%he previous example together with other not cited, shows
%results 
spanning the entire range of acceptable values: $0.2 \leq \Omega_{\rm m} \leq 1$ (see Reichart et al. 1999).
It is interesting to note that Viana \& Liddle (1999) using the same data set as Eke et al. (1998) showed that uncertainties both in fitting local data and in the theoretical modeling could significantly change the final results: they found $\Omega_{\rm m} \simeq 0.75$ as a preferred value with a critical density model acceptable at $<90\%$ c.l. while Eke et al. (1998) found $\Omega_{\rm m}=0.45 \pm 0.2$. 
%
%%Analyzing {\it EMSS}, Sadat, Blanchard
%%\& Oukbir (1998) and Reichart et al. (1999) found results consistent with $\Omega_{\rm m}=1$. A result consistent with $\Omega_{\rm m} \simeq 1$ %%was found by Blanchard \& Bartlett (1998), and Viana \& Liddle (1999) found that
%%$\Omega_{\rm m} \simeq 0.75$ with $\Omega_{\rm m}>0.3$ at the 90\% confidence level and $\Omega_{\rm m} \simeq 1$ still viable. Blanchard, %%Bartlett \& Sadat (1998) found almost identical results ($\Omega_{\rm m} \simeq 0.74$ with $0.3<\Omega_{\rm m}<1.2$ at the 95\% confidence level).
%%Eke et al. (1998) found $\Omega_{\rm m}=0.45 \pm 0.2$. 
%%It is interesting to note (as previously mentioned) that Viana \& Liddle (1999) used the same data set as Eke et al. (1998) and showed that %%uncertainties both in fitting local data and in the theoretical modeling could significantly change the final results: they found $\Omega_{\rm m} %%\simeq 0.75$ as a preferred value with a critical density model acceptable at $<90\%$ c.l.
%%Different results were obtained by Bahcall, Fan \& Cen (1997) ($\Omega_{\rm m}=0.3 \pm 0.1$), 
%%Bahcall \& Fan (1998) ($\Omega_{\rm m}=0.2^{+0.3}_{-0.1}$). The previous example together with other not cited, shows
%%results span the entire range of acceptable solutions: $0.2 \leq \Omega_{\rm m} \leq 1$ (see Reichart et al. 1999).
%

The reasons leading to the quoted discrepancies have been studied in several papers (Eke et al. 1998; Reichart et al. 1999; Donahue \& Voit 1999; Borgani et al. 2001) and can be summarized as due to: 
%According to Reichart (1999) unknown systematic effects may be plaguing great part of the quoted results. Systematic effects  entering the quoted %analysis are: 
1) The inadequate approximation given by the mass function used  (e.g., Bryan \& Norman 1998). 2) Inadequacy in the structure formation as described by the spherical model leading to changes in the threshold parameter $\delta_{\rm c}$ (e.g., Governato et al. 1999). 3) Inadequacy in the M-T relation obtained from the virial theorem (see Voit \& Donahue 1998; Del Popolo 2002). 
%\footnote{Even if Reichart et al. (1999), did some estimation of the changes produced by the previous systematic effects, as stressed by the same %authors a further investigation is needed taking into account the correct forms of the M-T relation and improved versions of the PS theory} 
4) Effects of cooling flows. 5) Determination of the X-ray cluster catalog's selection function.
5) Missing high redshift clusters in the data used (e.g., the {\it EMSS}). 6) Evolution of the L-T relation (Voit \& Donahue 1998).
7) The use of different best fitting procedures to get the constraints (Eke et al. 1998). 8) Other effects described in more recent papers (e.g., Mantz et al. 2008 (hereafter M08); Vikhlinin et al 2009b). 
%
%Voit \& Donahue (1999) point the attention on similar and different items to that stressed by Reichart et al. (1999):
%1) Deviation from the Press-Schechter orthodoxy (similarly to Reichart et al. 1999). 2) Missing high redshift clusters in the data used (e.g., the %{\it EMSS}). 3) Inadequacy in the M-T relation. 4) Evolution of the L-T relation.
%

%\end{document}

%%%%%%%%%%%%%%%%%%%%%%%%%%%%%%%%%%%%%%%%%%%%%%%%%%%%%%%%%%%%%%%%%%%%%%
%\end{document}
%%
%\frac{M}{\rho_m} \frac{n(M,z)}{d \ln \sigma^{-1}}
%\label{eq:jenk}
%\end{equation}    
%%

\begin{figure}[ht]
\centerline{\hbox{(a)
%\hspace{3.0cm} 
\psfig{file=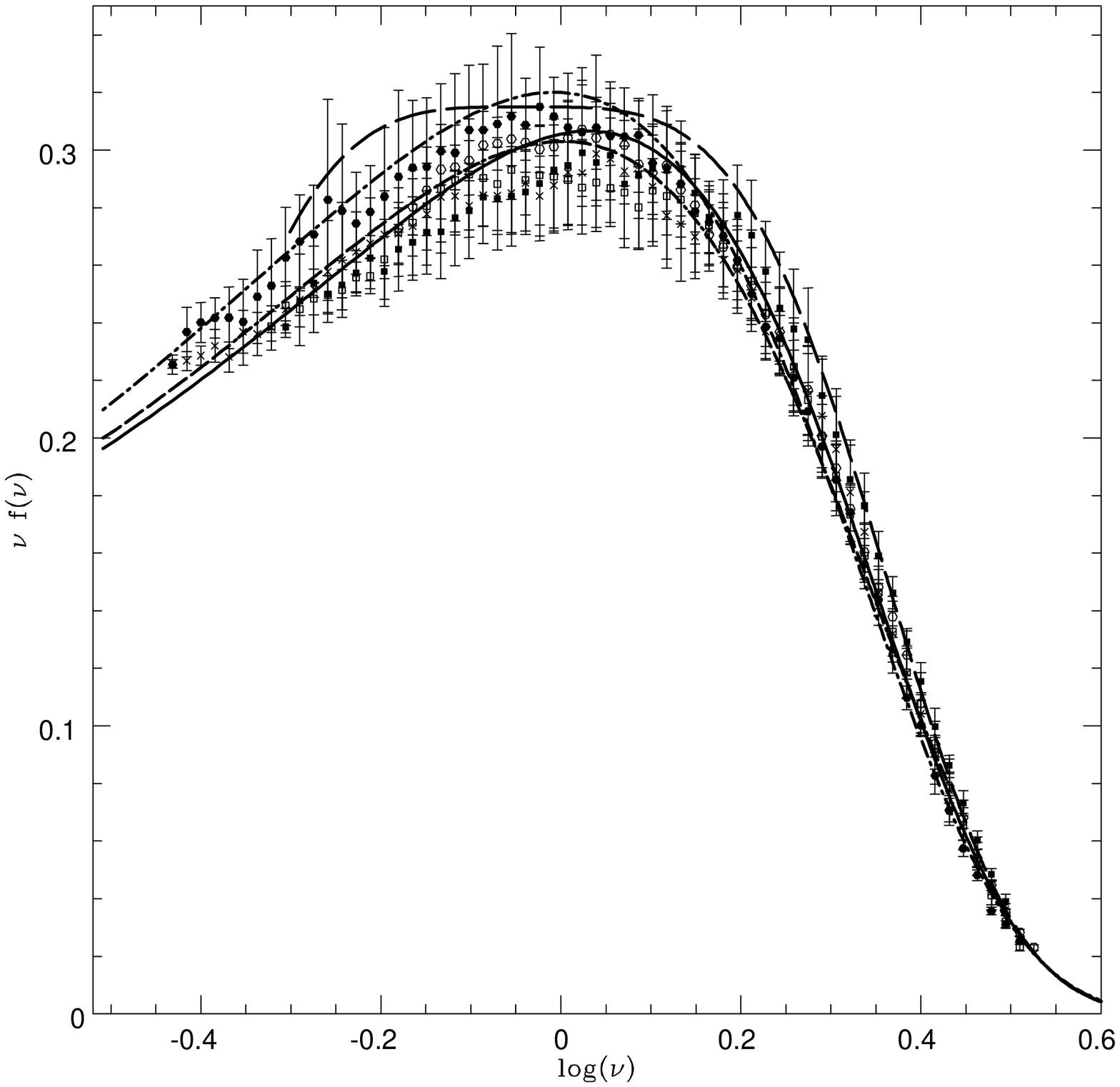,width=8.2cm} \hspace {0.1cm}(b)
}}
%\centerline{\hbox{(c)
%\psfig{file=tinkkk.eps,width=7cm}
%}}
\centerline{\hbox{(b)
\hspace{-1.2cm} 
\psfig{file=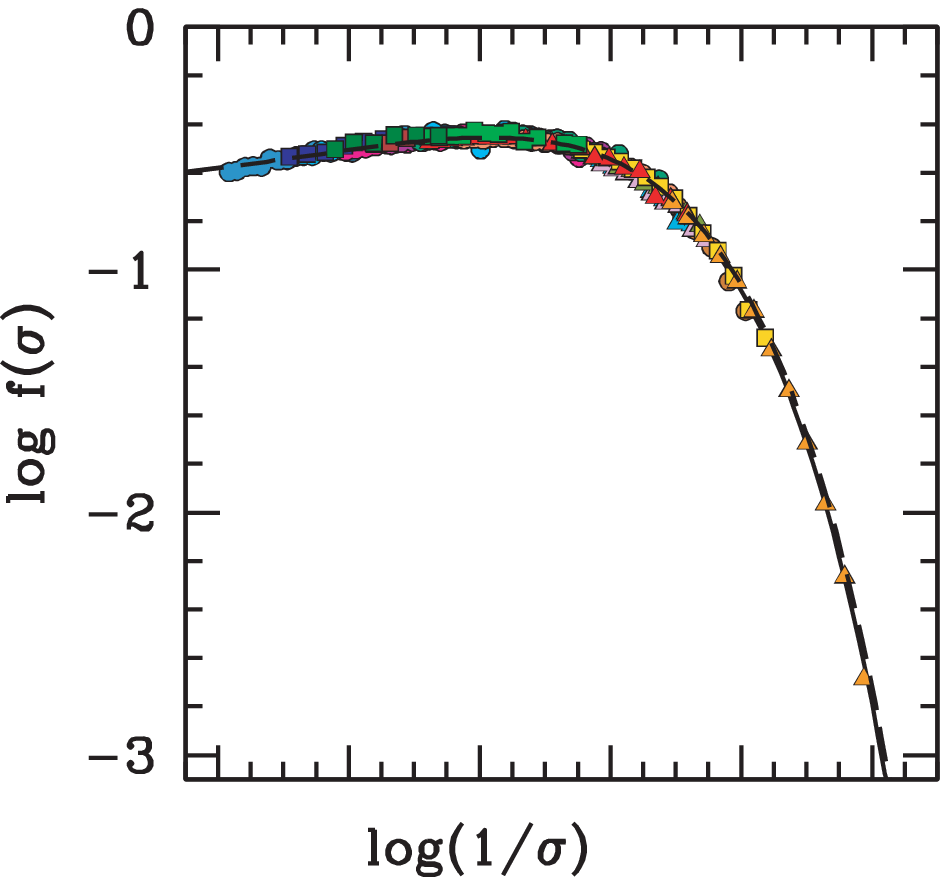,width=8.5cm} \hspace {0.1cm}(b)
}}
\caption{
In Panel (a), the solid line
represents the multiplicity function obtained in this paper, the short--dashed line the fitting formula proposed by Yahagi et al. (2004) (their Eq. 7), the dotted line the Sheth \& Tormen (2002) (ST) multiplicity function, the
long-dashed line the Jenkins et al. (2001) multiplicity function. The errorbars with
open circles represent the run 140 of YNY, those with filled squares
the case 70b, those with open squares the case 70a, those with filled
circles the case 35b, those with crosses the case 35a.
%Panel (b). Mass function plotted in redshift-independent form for all R03
%outputs:the redshifts used are 0,1,2,3,4,5,6.2,7.8,10,12.1, and 14.5.The
%solid line is the ST99 prediction, while the dashed and dotted lines represent the
%results of the present paper and eq. (16), respectively.
Panel (b). Mass function plotted in redshift-independent form. The measured $f (\sigma)$ from all simulations in Table 1 of Tinker et al. (2008). 
%%Results are presented at z = 0 and for ? = 200. 
The solid line is the best fit function of equation (3) (Tinker et al. 2008). The dashed line the model in the present paper.
}
\end{figure}
%
%%Eke et al. (1998), in a detailed study of the systematic uncertainties in the determination of $\Omega_{\rm m}$ showed that
%%even the differences between the derived best-fitting parameters from the $\chi^2$ and the Maximum likelihood techniques has non-negligible %%contributions in the systematic uncertainties.
%%As stressed by Henry (2000), the maximum likelihood fit only provides the best fit, it does not provide an assessment of whether that
%%fit is a good fit to the data. Marshall et al (1983) require that the two independent cumulants (see Eqs. 7, 8 of
%%Henry (2000)) both be uniform according to the two­-tailed Kolmogorov­-Smirnov test. 
%%They recommend rejecting the
%%model if the probability of either being uniform is less than 0.05.
%
%
%%Although the quoted uncertainties has been so far of minor importance with respect to the paucity of observational data, a breakthrough is needed %%in the quality of the theoretical framework if high-redshift clusters are to take part in the high-precision-era of observational cosmology.
%
%In recent years, the constraints have been improved thanks 
The situation with the cluster mass function data has been
dramatically improved in the past years. A large sample
of sufficiently massive clusters extending to $z \simeq 0.9$ has been
derived from ROSAT PSPC pointed data covering 400 deg$^2$
(Burenin et al. 2007). Distant clusters from
the 400d sample were then observed with Chandra, providing
high-quality X-ray data and much more accurate total mass
indicators (see Vikhlinin et al. 2009b). Chandra coverage has also become available for
a complete sample of low-z clusters originally derived from
the ROSAT All-Sky Survey (see Vikhlinin et al. 2009b). 
Results from deep Chandra pointings
to a number of low-z clusters have significantly improved
our knowledge of the outer cluster regions and provided a
much more reliable calibration of the $M_{tot}$ vs. proxy relations
than what was possible before. 
On the theoretical side, improved
numerical simulations resulted in better understanding
of measurement biases in the X-ray data analysis (Nagai
et al. 2007; Rasia et al. 2006; Jeltema et al. 2007). 
%Even more
%importantly, results from these simulations have been used to
%suggest new, more reliable X-ray proxies for the total mass
%(Kravtsov et al. 2006). We discuss all this issues in the previous
%paper (Vikhlinin et al. 2008, Paper II herea.er). e cluster
%mass functions derived in this paper are reproduced in Fig. 1.
%Overall, these results are an important step forward in providing
%observational foundation for cosmological work with the
%cluster mass functions.
%= - ± evolving w in .at universe (§ 9.1) and constant w in non-.at
%universe (§ 9.2)

In the present paper, we want to show how tighter constraints
can be obtained in M08 model improving the mass function adopted by them, and the scaling laws used (e.g., the M-T and M-L relationships). 
%%%%%%%%%%%%....................
In this paper, we use the observed X-ray luminosity
function to investigate two cosmological
scenarios, assuming a spatially flat metric
in both cases: the first includes dark energy in
the form of a cosmological constant ($\Lambda$CDM);
the second has dark energy with a constant
equation-of-state parameter, w (wCDM). The
theoretical background for this work is reviewed
in Section 2. Section 3 presents the results
and Section 4 the conclusions.

\section{Theory}

In the introduction, we discussed the ingredients needed in the recipe used to constrain
cosmological parameters from X-ray observations. In this section, I derive an expression for
the X-Ray luminosity function (XLF) (using now the mass function obtained in Del Popolo (2006a, b)) and M--T, L--T relations obtained in Del Popolo 2002,
and Del Popolo et al. 2005, respectively) and then I set some constraints to 
$\Omega_m$, $\sigma_8$ and the dark--energy equation--of--state parameter $w$, by using the data (clusters) used in M08, namely MACS (Massive Cluster Survey), BCS (Brightest Cluster Sample), and REFLEX (ROSAT ESO FLUX LIMITED X-Ray SAMPLE).
Following M08, the constraints are obtained from measurements of the X-Ray luminosity function of the quoted samples. 
The most straightforward mass–-observable relation to complement these X--ray flux--limited surveys is the mass–-X--ray luminosity relation. For sufficiently massive (hot) objects at the relevant redshifts, the conversion from X-ray flux to luminosity is approximately independent of temperature, in which case the luminosities can be estimated directly from the survey flux and the selection function is
identical to the requirement of detection. 
%In a flux-complete survey further restricted to high luminosities, every cluster should thus be usable in the analysis, without the need for
%additional observations other than those required to calibrate the mass–luminosity relation. 
A disadvantage is that there is a large scatter in cluster luminosities at fixed mass; however, sufficient data allow this scatter to be quantified
empirically. More recently, a dramatic reduction in luminosity-mass scatter has been demonstrated when luminosities are measured excluding 
cluster centers (typically $r<0.15 r_{500}$; Maughan 2007; Zhang et al. 2007).
Alternative approaches use cluster temperature (Henry 2000; Seljak 2002; Pierpaoli et al. 2003; Henry 2004),
gas fraction (Voevodkin \& Vikhlinin 2004) or $Y_X$ parameter (Kravtsov et al. 2006) to achieve tighter mass–-observable relations at the expense of reducing the size of the samples available for analysis. The need to quantify the selection
function in terms of both X-ray flux and a second observable additionally complicates these efforts.
 
The first ingredient of the quoted recipe (i.e., mass function), used in M08 was the Jenkins et al. (2001)  (hereafter J01) mass function. J01 wrote the mass function of galaxy clusters of mass $M$ at redshift $z$ as a "universal function" of $\sigma^{-1} (M,z)$
\begin{equation}
f(\sigma^{-1})= \frac{M}{\rho_m} \frac{n(M,z)}{d \ln \sigma^{-1}}
\label{eq:jenk}
\end{equation}    
which was fitted by
\begin{equation}
f(\sigma^{-1})= A e^{(-|\ln \sigma^{-1}+B|^{\epsilon})}
\end{equation}  
for cosmological -constant models, with $A=0.316$, $B=0.67$, and $\epsilon=3.82$.

As shown in Del Popolo (2006a) and Del Popolo (2006b), the theoretical mass function obtained in the quoted papers is in better agreement with high
resolution N-body simulations, namely Reed et al. 2003 (R03), Yahagi et al. (2004) (YNY), Warren et al. 2006 (W06), and Tinker et al. (2008) (see the following and Fig. 1b) (T08).   %(BIBLIOGRAFIA). 

The mass function was calculated according to the model of Del Popolo (2006a, b).   
%in agreement with N-body simulations like those of Yahagi, Nagashima \& Yoshii 2004)
%
%; Warren  et al. 2005 (W05)). 
%, and Engineer et al. 2000).
%}
The multiplicity function, in the quoted model, is given by:
\begin{eqnarray}
\nu f(\nu )&=& A _1 \left( 1+\frac{\beta_1 g(\alpha_1)}{\left( a\nu \right) ^{\alpha_1}}
+\frac{\beta_2 g(\alpha_2)}{\left( a\nu \right) ^{\alpha_2}}+\frac{\beta_3 g(\alpha_3)}{\left( a\nu \right) ^{\alpha_3}}
\right) 
\nonumber \\
& &
\sqrt{\frac{a\nu }{2\pi }}
e^{\{\frac{-a \nu}{2} \left[ 1+\frac{\beta_1}{\left( a\nu \right) ^{\alpha_1}}
+\frac{\beta_2}{\left( a\nu \right) ^{\alpha_2}}+\frac{\beta_3}{\left( a\nu \right) ^{\alpha_3}}
\right] ^{2}\}}
%\right] ^{2}/2\}}
\label{eq:mia}
\end{eqnarray}
where
\begin{eqnarray}
g(\alpha_i)&=&
\mid 1-\alpha_i +\frac{\alpha_i (\alpha_i
-1)}{2!}-...- \nonumber\\
& &
\frac{\alpha_i(\alpha_i-1)\cdot \cdot \cdot
(\alpha_i-4)}{5!} \mid
\label{eq:miaa}
\end{eqnarray}
where $i=1$ or 2, $\alpha_1=0.585$, $\beta_1=0.46$, $\alpha_2=0.5$ and $\beta_2=0.35$, $\alpha_3=0.4$ and $\beta_3=0.02$, $a=0.707$, and $A_1=1.2$ is
the normalization constant.

The ``multiplicity function" is correlated with the usual, more straightforwardly used, ``mass function" as follows.  
Following Sheth \& Tormen (2002) (hereafter ST) notation, if $f(M,\delta) dM$ denotes the fraction of mass that is contained in collapsed haloes that have mass in the range $M$-$M+dM$, at redshift $z$, and $\delta(z)$ is the redshift dependent overdensity, 
the associated ``unconditional" mass function is:
\begin{equation}
n(M,\delta)dM=\frac{{\rho_b}}{M} f(M,\delta) dM
\label{eq:mfu}
\end{equation}

In Fig. 1a, we plot the multiplicity function obtained in this paper (simbols are described in the figure caption).
%the solid line represents the multiplicity function obtained in this paper, the short--dashed line the fitting formula proposed by YNY (their Eq. 7), the dotted line the ST %multiplicity function, the long-dashed line the J01 multiplicity function. The errorbars with open circles represent the run 140 of YNY, those with filled squares
%the case 70b, those with open squares the case 70a, those with filled circles the case 35b, those with crosses the case 35a.

%{\bf 
There are some differences
%discrepancies 
between the quoted simulations 
%(agreeing with Del popolo 2006a,b model) 
and the J01 simulations. 
First, the multiplicity function of the present paper, similar to that of YNY, in
the low-$\nu$ region of $\nu \leq 1$ systematically falls below the J01 functions. In this region the multiplicity function of the
present paper is very close to that of YNY.
%As seen in Fig. 1, and in agreement with YNY, the numerical
%multiplicity functions reside between the ST and J01 multiplicity
%functions at 2 ¡Ü í ¡Ü 3 (except for the run 35b).
Additionally, the numerical multiplicity functions (and that in Del Popolo 2006a,b) have an apparent
peak at $\nu \simeq$ instead of the plateau that is seen in the
J01 function.
Similar differences are seen in the high-$\nu$ region.
%, where $\nu$ is significantly
%larger than unity, the multiplicity function of the
%present paper like YNY takes values between ST and J01 functions.
These differences between numerical multiplicity functions (R03; YNY; W05; Del Popolo 2006a,b)
and J01, are however within 1--$\sigma$ error--bars, and so they are overall in agreement.
%, and they are possibly due to the di.erent box
%sizes adopted (see YNY for a discussion). 
%
%%Throughout the peak range of $0.3 \leq \nu \leq 3$, the ST multiplicity function is in disagreement
%%with the high mass resolution N-body simulations
%%of YNY and that of the present paper. 
%
%As shown by YNY the
%ST functional form provides a good fit to them only choosing
%parameter values of a = 0.664, p = 0.321, and A3 = 0.301. 
The multiplicity function obtained in the present paper has a peak
at $\nu \simeq 1$ as in YNY numerical multiplicity function, instead of a plateau as in the
J01 function. Differences are observed also in the redshift evolution of the J01 mass function (Del Popolo 2006b).
Summarizing the fitting formulas presented by J01 are accurate to $\simeq 10- 20 \%$ (Tinker et al. 2008 (T08)). 
In our model, 
%that we used, 
the mass function that we used is given by Eq. (\ref{eq:mia}), Eq.(\ref{eq:miaa}), and Eq(\ref{eq:mfu})  
%similarly to W05 and T08 
which is in perfect agreement with the T08 mass function, as shown in Fig. (1b).
So, the accuracy of the mass function is, as in T08, of the order of $\simeq 5 \%$ for $\Lambda$CDM models for the 
mass and redshift range of interest in this study. As a consequence, in this way the theoretical uncertainties in the mass function do not contribute
significantly to the systematic error budget. 

In Fig. 1b, I plot the mass function for all of our outputs in the $f(\sigma)- \ln(\sigma^{-1})$ plane.  Large values of $\ln\sigma^{-1}$
correspond to rare haloes of high redshift and/or high  mass, while
small values of $\ln\sigma^{-1}$ describe haloes of low mass and
redshift  combinations.  
%In Fig. 2, I plot the evolution of the mass function over all of our redshifts as a function of $M/M_*$.  
%$M_*$ ?????
% 
%%The solid line is the ST mass function while the dashed line the one obtained in the present paper, the dotted line represents 
%%Eq. (9) in R03 (improvements to ST model), and data are from Ro3 simulations. The ST and the mass function of the present paper differs more in %%the high mass region, where the mass function of the present paper is steeper than ST
%%and in better agreement with numerical simulations data than ST mass function. 
%%The ST function fits the simulated mass function to better than 10$\%$ over the range of -1.7 $\leq
%%\ln\sigma^{-1} \leq$ 0.5 while it  
%%appears to significantly overpredict haloes for $\ln\sigma^{-1} \geq$ 0.5.  
%%YNY and Warren et al. (2005) made a similar choice, namely they introduced an empirical mass function obtained from a fit to their simulations 
%%that gives a better fit to simulations than ST model.
%
Fig. 1b shows the function $f(\sigma)$ measured for all simulations in Table 1 of T08, the solid line the fit to the data (namely T08 eq. 3) and the dashed line the model of the present paper.

As previously reported, one of the main problems of using the mass function to constrain cosmological parameter is that theoretical 
predictions provide the number density of clusters of a given mass, while the mass itself is never the directly observed quantity. 
One then needs relations connecting mass with other quantities more easily obtainable which can be used as a surrogate for cluster mass.
Over the past decade, observations of clusters of galaxies (e.g. ROSAT, ASCA) have shown the existence of a 
%tight 
correlation 
between the total gravitating mass of clusters, $M_{tot}$\footnote{Since $M_{tot}$ compares with the ICM temperature measurements that can be obtained through X-ray spectroscopy, this explains the importance of a mass–-temperature (M–T) relation.}, their X-ray luminosity ($L_X$) and the temperature ($T_X$) of the intracluster medium (ICM) (David et al. 1993; Markevitch 1998; Horner, Mushotzky \& Scharf 1999).  
By means of the quoted scaling relations one can obtain 
%There are 
different methods for tracing the evolution of the cluster number density: (1) The X-ray temperature
function (XTF), which has been presented for local (e.g., Henry \& Arnaud 1991) and distant clusters (Eke et al. 1998; Henry
2000). 2) The evolution of the X-ray luminosity function (XLF). In this case, we need a relation between the observed $L_X$ and the cluster virial mass. 
%
%%On one hand, the X-ray temperature measures the depth of the potential wells, and the bolometric luminosity, $L \propto n^2 R^3_X T^{1/2}$, %%emitted as thermal bremsstrahlung by intracluster plasma measures the baryon number density n within the volume
%%$R^3_X$. By means of the quoted scaling relations one can obtain 
%There are 
%%different methods for tracing the evolution of the cluster number density: (1) The X-ray temperature
%%function (XTF), which has been presented for local (e.g., Henry \& Arnaud 1991) and distant clusters (Eke et al. 1998; Henry
%%2000). 2) The evolution of the X-ray luminosity function (XLF). In this case, we need a relation between the observed $L_X$ and the cluster virial %%mass. 
%{}
%

In the following, following M08, we shall use the XLF to constrain cosmological parameters. 
Then the next crucial step, after having a mass function, is to convert it in a Luminosity function (XLF). This 
can be done by first converting mass into intra-cluster gas temperature, by means of the $M-T_{x}$ relation, and then converting the temperature into X-ray luminosity, by means of the $L_{x}-T_{x}$ relation. 
M08 used a self-similar relationship between mass and X-ray luminosity for massive clusters (e.g., Bryan \& Norman 1998) modified by an additional redshift--dependent factor (see Morandi et al. 2007).  
At this point, we must stress an important point. Numerical simulations confirm that the DM component in clusters of galaxies, which represents the dominant fraction of the mass,
has a remarkably self-similar behavior; however the baryonic component does not show the same level of self-similarity. This picture
is confirmed by X-ray observations, see for instance the deviation of the L–-T relation in clusters, which is steeper than the theoretical
value predicted by the previous scenario.
More precisely, until some years ago, the cluster structure was considered to be scale-free, which means that the global properties of clusters,
such as halo mass, luminosity-temperature, and X-ray luminosity would scale self-similarly (Kaiser 1986; Evrard \& Henry 1991). 
In particular, the gas temperature would scale with cluster mass as $T \propto M^{2/3}$ and the
bolometric X-ray luminosity would scale with temperature as $L \propto T^{2}$, in the bremsstrahlung-dominated regime above $2$ keV.
Studies following that of Kaiser (1986) showed that the observed luminosity-temperature relation is closer to $L \propto T^{3}$
(e.g., Edge \& Stewart 1991), indicating that non--gravitational processes should influence the density structure of a cluster's core,
where most of the luminosity is generated (Kaiser 1991; Evrard \& Henry 1991; Navarro et al. 1995; Bryan \& Norman 1998). One way to obtain a scaling law closer to the observational one is to have non--gravitational energy injected into the ICM before or during cluster formation, the so-called pre-heating (Ponman et al. 1999; Bower et al. 1997; Cavaliere et al. 1997, 1999; Tozzi \& Norman 2001; Borgani et al. 2001; Voit \& Brian 2001), feedback processes that alter the gas characteristics during the evolution of the cluster (Voit \& Bryan 2001), cooling flows (Allen \& Fabian
1998).
A similar situation is valid for the M-T relationship,
%recent studies have shown 
namely that the self-similarity in the
M–-T relation seems to break at a few keV (Nevalanien et al. 2000; Xu, 
%Jing \& Wu 2001; Xu et al. 2001; 
Finoguenov et al. 2001; Muanwong et al. 2001; Bialek, Evrard \& Mohr 2001).
%; afshordi \& Cen.....). 
Consequently, if, as in M08, one starts with self-similar scaling laws in order to have consistent scaling relations one has to compare the self-similar scaling relations to observations (Morandi et al. 2007).
%, which allows us to evaluate the importance of the effects of the non-gravitational processes on the ICM physics (see Morandi et al. 2007).

Different from the M08 approach, in the following, we use models for the L--T, T--M, relationships taking into account the non-self similarity: namely, the $M-T_{x}$ relation obtained analytically using the model of Del Popolo (2002),  
%which takes into account 
%from the continuous formation model based on the fact that clusters form gradually, is taken into account by 
%means of the merging-halo formalism (Lacey \& Cole 1993)
while the $L_{x}-T_{x}$ relation is that obtained in Del Popolo, Hiotelis \& Pe\'narrubia (2005) based on an improvement of the  
Punctuated Equilibrium Model (PEM) of Cavaliere et al. (1997, 1998, 1999). 
%The reason, we used this approach instead of that of Mantz et al. (2008) was already described in Del Popolo (2003) and in the following.
The drawbacks of using self-similar relationships fitted to the data (clusters) and the reasons to use a different approach, were already discussed in Del Popolo (2003) (their sect. 3), and in the remainder of this section.
% (ApJ 599, 723).

Similarly, to the present study, in Del Popolo (2003) we used the models for the L--T, T--M, relationships instead of the scaling relations
obtained from simulations of Chandra data (see, e.g., Pierpaoli et al. 2001, 2003 for references).\footnote{Notice that Eq. 22 in Del Popolo 2003, and Eq. 5 in M08, are very similar.}
Eq. (5) in M08 similar to that Eq. (13) of Pierpaoli et al. (2001) or Eq. (4) of Pierpaoli
et al. (2003) comes from rather simplistic arguments (dimensional analysis and an assumption that clusters are self-similar)
not taking into account important physical effects that gives rise to a non-self-similar behavior of the quoted relation, as previously 
discussed. The fitting procedure used by M08, trying to take account of the previous physics and the non-self-similar behavior of the relationship, is complicated by several effects.
%
%%and is a {\bf good approximation to both observations and simulations but are sufficiently computationally demanding that they
%%cannot explore parameter space efficiently, and so it is necessary to determine coefficients by means of simulations, while
%%scalings are taken from simple theoretical models (Pierpaoli et al. 2001)}. 
%
In fact, in the fit 
%in Mantz et al. (2008) Eq.  (7),(8) 
one uses data that may contain small groups which can be influential in the estimation of the slope of the model, and  
%Since we
%only require the mass–luminosity relation 
one has then to choose accurately the data to be used in the fit. 
This choice mitigates the possibility of obtaining biased results if
slope of the mass–-luminosity relation is different for massive
clusters compared with smaller groups.
In M08 they fitted only the data (clusters) with 
%$L> 3 \times 10^{44} E(z) $ erg/s  (E(Z)?????).  
$L> 3 \times 10^{44}$ erg/s in the 0.1--2.4 keV band.  
Moreover, the process of fitting the model in Eq. 7 of M08 is complicated by the presence of Malmquist bias. 
Close to the flux limit for selection, any X-ray selected sample will preferentially include the most luminous sources for a given mass.
This results in a steepening of the derived mass–-luminosity relation and a bias in the inferred intrinsic scatter in luminosity for a given mass.
The use of the extended sample of Reiprich \& B\"oringer (2002) (RB02), rather than only their flux--limited HIFLUGCS sample, partially mitigates this effect by softening the flux limit.
A further problem is that as a consequence of Malmquist bias there is a strong apparent, but not necessarily physical, correlation between
luminosity and redshift due to the fact that the flux limit corresponds to higher luminosities at higher redshifts. \\
%
%\footnote{In PEM CMT98), the
%cluster evolution is described as a sequence of ''punctuated equilibria,''
%that is to say, a sequence of hierarchical merging episodes
%of the DM halos, associated in the intracluster plasma ( ICP) with
%shocks of various strengths (depending on the mass ratio of the
%merging clumps), which provide the boundary conditions for the
%ICP to readjust to a new hydrostatic equilibrium.
%}
%  

%  TOLTO
%%%%%In order to check the validity of our approach and the L--M relationship that we used, we also performed an analysis similar to that of M08, %%%%%described in their Sect. 3.1. Namely we performed a similar fit to the RB02 sample, as done in M08, but using our L--M relationship, and %%%%%assuring that our L--M relationship was at least as good as that used by M08. 
%

%%COMUNQUE ABIAMO VERIFICATO CHE IL NOSTRO RISULTATO FOSSE ALMENO UGUALE A QUELLO DI MANTZ, OPPURE FATTO UN FIT USANDO I NOSTRI MODELLI ANALITICI.\\

For what concerns the data (clusters) used in the analysis, they are the same of those used by M08: the following 
three flux-limited surveys are included in our analysis: the BCS (Ebeling et al. 1998) and REFLEX sample (B\"ohringer et al. 2004) at low redshifts ($z < 0.3$), and the MACS (Ebeling et al. 2001) at $0.3 < z < 0.5$ (see M08). In the analysis, the sample was chosen to cover the redshift range $z<0.5$, since at higher redshifts the number of unrelaxed clusters decrease, and the L--T and T--M relations are appropriate for relaxed clusters.
The purpose of this paper is to present an analysis based only on the X-ray luminosity function (XLF) data described above, along with the priors described in Sect. 4 of M08.

Following M08, we parametrize the full model fitted to the X-ray luminosity function data as 
{$h$, $\Omega_b h^2$, $\Omega_c h^2$, $\sigma_8$, $n_s$, $w$}
%, $\alpha$, $\beta$, $\gamma$, $\eta_0$, $\eta_1$, $B$}, 
%%%%%%%%%%%%%%%%%%%%%%%%%%%%%%%%%%%%%%%%%%%%%%%%%%%%%%%%%%%%%%%%%%%%SISTEMARE
where $\Omega_b$ and $\Omega_c$ are the baryon and cold dark matter densities ($\Omega_m=\Omega_b+\Omega_c$).
%
%$A$ is the normalization of the Jenkins mass function (\eqnref{eq:jmf}), 
%%%%%%%%%%%%%%%%%%%%%%%%%%%%%%%%%%%%%%%%%%%%%%%%%%%%%%%%and the last 6 parameters describe the mass--luminosity relation and its calibration. 
In addition to the assumption of spatial flatness, we adopt the Gaussian priors $h=0.72 \pm 0.08$ (Freedman et al. 2001) and $\Omega_b h^2=0.0214 \pm 0.002$ (Kirkman et al. 2003) from the Hubble Key Project and Big Bang nucleosynthesis studies, respectively. 
%
%%These latter priors are necessary because the likelihood depends very weakly on $h$ and $\Omegab$, which enter only through the transfer function. %%We additionally marginalize over a 20 per cent uncertainty in the normalization, $A$, of the \JMF{} mass function to account for the residuals of %%the fitting formula to their simulations over the mass range of interest and the expected variation among cosmologies. We must also place a prior %%on the mass--luminosity evolution parameters $\gamma$ and \iscatprime{}, which are not constrained by our analysis of the \RB{} data. 
%%For the standard set of allowances used in this paper, we adopt the uniform priors $|\gamma|<0.35$, which corresponds to a limit of approximately %%20 per cent evolution in the normalization of the mass--luminosity relation within redshift 0.7, and $|\iscatprime|<0.3$. 
Since the results are insensitive to the spectral index within a reasonable range (see M08), we fix $n_s=0.95$ in accordance with (Spergel et al. 2007) for the standard analysis. The dark-energy equation of state was bounded by a uniform prior, $-5<w<0$. 
%%For the remaining model parameters, uniform priors on the physically allowed domains were used. These priors are summarized in Tab. 1 of mantz et %%al. (2008). \tabref{tab:priors} (labeled ``standard'' priors). The sensitivity of our results to the choice of priors is analyzed in \secref{sec:priors}.
%
%%%{it\ 
%%%Cosmological constraints were determined via Markov Chain Monte Carlo, employing the Metropolis algorithm; the calculations were performed using %%%a{ modified version of the COSMOMC code\footnote{http://cosmologist.info/cosmomc/} of \citet[][see also \citealt{Rapetti05,Rapetti07}]{Lewis02}. 
%%%it Multiple, randomly initialized Markov chains were produced for each set of results, and convergence to the posterior distribution was %%%monitored using the Gelman-Rubin criterion, $R$, which measures the ratio of between-chain to within-chain variances \citep{Gelman92}, as well as %%%by visual inspection. Acceptable convergence was defined by the requirement $R-1<0.05$. The log-likelihood of the data given a set of model %%%parameters is decomposed into the sum $\ell_{\alpha\beta} + \ell_\iscatsym + \llxlf$, whose terms are described in the remainder of this Section. }
%

The luminosity function likelihood is the same as in M08 (Sect. 4.2).
% recalling that in their Eq. (16)  

The likelihood that $N$ clusters with inferred luminosities in a range $d\hat{L}$ exist in a volume $dV$ can in general be written as a Poisson probability plus a correction due to the clustering of halos with one another. 
If the plane of redshift and inferred luminosity is divided into non-overlapping cells, then the likelihood of our data is simply
\begin{equation}  \label{eq:xlflike}
  P\left(N_1,N_2,\ldots\right) = \prod_j \frac{\tilde{N}_j^{N_j}e^{-\tilde{N}_j}}{N_j!},
\end{equation}
where $N_j$ and $\tilde{N}_j$ are the number of clusters detected and predicted in the $j$th cell, respectively. 
%The log-likelihood is then
%\begin{equation}
%  \label{eq:xlfloglike}
%  \llxlf = \sum_j \left[ N_j\ln(\tilde{N}_j) - \tilde{N}_j \right],
%\end{equation}
%where the model-independent term $-\sum_j \ln(N_j!)$ has been dropped.

If the cells are taken to be rectangular, with the $j$th cell given by $z_j^{(1)} \leq z < z_j^{(2)}$ and $\hat{L}_{j}^{(1)} \leq \hat{L} < \hat{L}_{j}^{(2)}$, then
\begin{equation}
  \label{eq:modelN}
  \tilde{N}_j = \int_{z_j^{(1)}}^{z_j^{(2)}}dz \frac{dV(z)}{dz} \int_{\hat{L}_{j}^{(1)}}^{\hat{L}_{j}^{(2)}}d\hat{L} \frac{d\tilde{n}(z,\hat{L})}{d\hat{L}},
\end{equation}
where $V(z)$ is the comoving volume within redshift $z$. In the absence of intrinsic scatter in the mass--luminosity relation and measurement errors in the observed luminosities, the derivative of the comoving number density would be simply
\begin{equation}
  \label{eq:dndLsimple}
  \frac{d \tilde{n}(z,L)}{dL} = f_{sky}(z,L) \frac{dM(L)}{dL} \frac{dn(z,M)}{dM} 
%\frac{dM_{324m}}{dM_{500}}.
\end{equation}
Here $f_{sky}$ is the sky coverage fraction of the surveys as a function of redshift and inferred luminosity, 
$dn/dM$ is no longer the Jenkins mass function but the one discussed in the present paper 
%(\eqnref{eq:mfcn_def}) 
and $M(L)$ is the mass--luminosity relation discussed in the present paper. 
%(\eqnref{eq:MLpowlaw}). The presence of scatter requires us to take into account that a cluster detected with inferred luminosity $\hat{L}$ could %potentially have any true luminosity $L$ and mass $M_{500}$, with some associated probability.

Similar to M08, the presence of scatter requires us to take into account that a cluster detected with inferred luminosity $\hat{L}$ could potentially have any true luminosity $L$ and mass $M$, with some associated probability. To calculate the predicted number density correctly, we must therefore convolve with these probability distributions:
\begin{eqnarray}
  \label{eq:dndL}
  \frac{d\tilde{n}(z,\hat{L})}{d\hat{L}} & = & f_{sky}(z,\hat{L}) \int_0^\infty dL ~ P(\hat{L}|L) \\ 
  & \times & \int_0^\infty dM ~ P(L|M) \frac{dn(z,M)}{dM} 
%\frac{dM_{324m}}{dM_{500}}
.\nonumber
\end{eqnarray}
$P(L|M)$ is a log-normal distribution whose width is like in M08 the intrinsic scatter in the mass--luminosity relation, 
$\eta(z)$, and $P(\hat{L}|L)$ is a normal distribution whose width as a function of flux is modeled as a power law, as described in Sect. 3.2 of M08. 
%
%%When evaluating the conversion between $M_{500}$ and $M_{324m}$ using the method of Hu \& Kravtsov (2003), we assume a universal concentration of %%4.0 at radius $r_{200}$, consistent with the results of Gao et al. (2007), who find that the concentrations of very massive clusters varies little %%with mass and redshift through $z=1$. 
%
%We have verified that including a log-normal scatter in concentration of $\sim 0.15$ \citep[e.g.][]{Comerford07,Schmidt07} has no effect on our %results.

\section{Results}

In Fig. 2, we compare, for a $\Lambda$CDM the joint $\Omega_m$-$\sigma_8$ constraints obtained from the BCS, REFLEX and MACS data sets combination 
%individually, as well as their combination,  
The marginalized constraints from the combination of the three cluster samples are $\Omega_m=0.28^{+0.05}_{-0.04}$ and $\sigma_8=0.78^{+0.04}_{-0.05}$
while the result of M08 gives less tight constraints $\Omega_m=0.28^{+0.11}_{-0.07}$ and $\sigma_8=0.78^{+0.11}_{-0.13}$.

\begin{figure}[ht] 
\centerline{\hbox{
\psfig{file=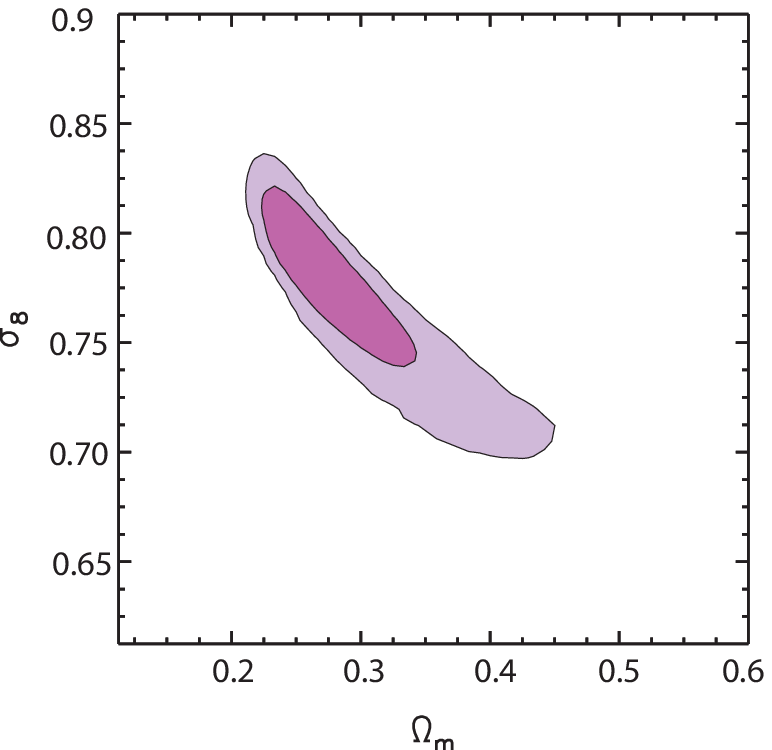,width=7cm}
}} 
\caption{
Joint 68.3 and 95.4 per cent confidence constraints on $\Omega_m$ and $\sigma_8$ for a $\Lambda$CDM model from MACS, BCS, and REFLEX combination using standard priors (as described in the text). 
%%The 68\% 
%(inner region) 
%%and 95\% 
%(outer region) 
%%credible regions of the marginalized posterior probability distribution P($\sigma_8$,$\Omega_m$) for a $\Lambda$CDM  model. 
}
\end{figure}

%%%%%%%Comparing this results with the similar one of M08, 
%$\Omega_m=0.28^{+0.11}_{-0.07}$ and $\sigma_8=0.78^{+0.11}_{-0.13}$,
%%%%%%%we see a noteworthy improvement in the constraints. 

%{\bf 
Our previous constraints are in good agreement with recent, independent results from the CMB (Spergel et al. 2007) and cosmic shear, as measured in the 100 Square Degree Survey (Benjamin et al. 2007) and CFHTLS Wide field (Fu et al. 2008). Our results are also in good overall agreement with previous findings based on the observed X-ray luminosity and temperature functions of clusters (Eke et al 1998, Donahue \& Voit 1999, Henry 2000, Borgani et al. 2001, Seljak 2002, Allen et al. 2003, Pierpaoli et al. 2003, Schuecker et al. 2003, Henry 2004).
%, although the correction to the hydrostatic mass estimates employed in the present study leads to our result on $\sigma_8$ being, typically, %somewhat higher for a given value of $\Omega_m$. 
Our result on $\Omega_m$ is in excellent agreement with current constraints based on cluster $f_{gas}$ data (Allen et al. 2008 and references therein) and the power spectrum of galaxies in the 2dF galaxy redshift survey (Cole et al. 2005) and Sloan Digital Sky Survey (SDSS) (Eisenstein et al. 2005, Tegmark et al. 2006, Percival et al. 2007), as well as the combination of CMB data with a variety of external constraints (Spergel 2007). 
%Our result on  $\sigma_8$ is marginally lower than that determined by weak lensing tomography in the Cosmic Evolution Survey (COSMOS; Massey07) and %by the observed number density of optically-selected groups and clusters in the 2dF (Eke06) and SDSS surveys (Rozo07).
%}

\begin{figure}[ht] 
\centerline{\hbox{
\psfig{file=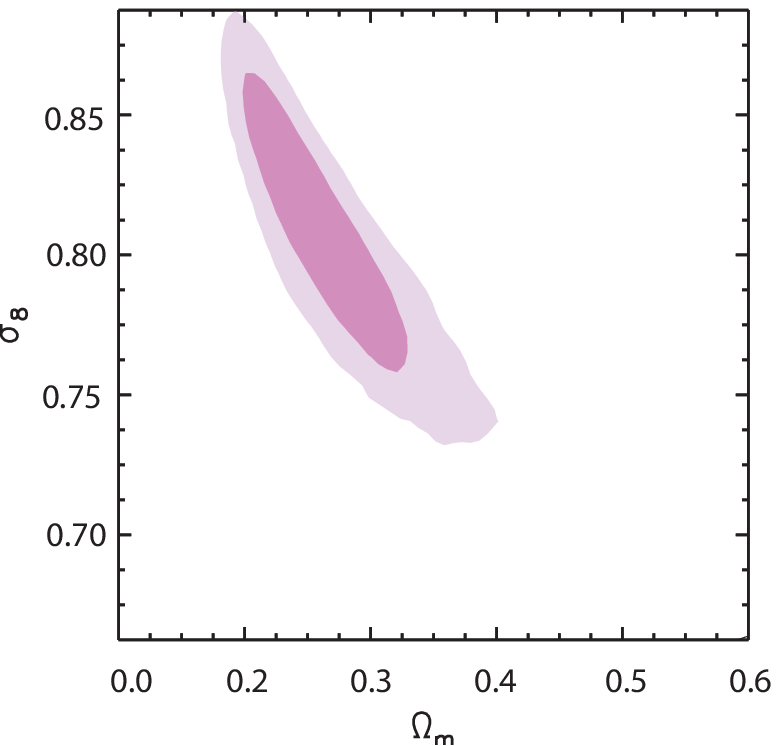,width=4.5cm}
\psfig{file=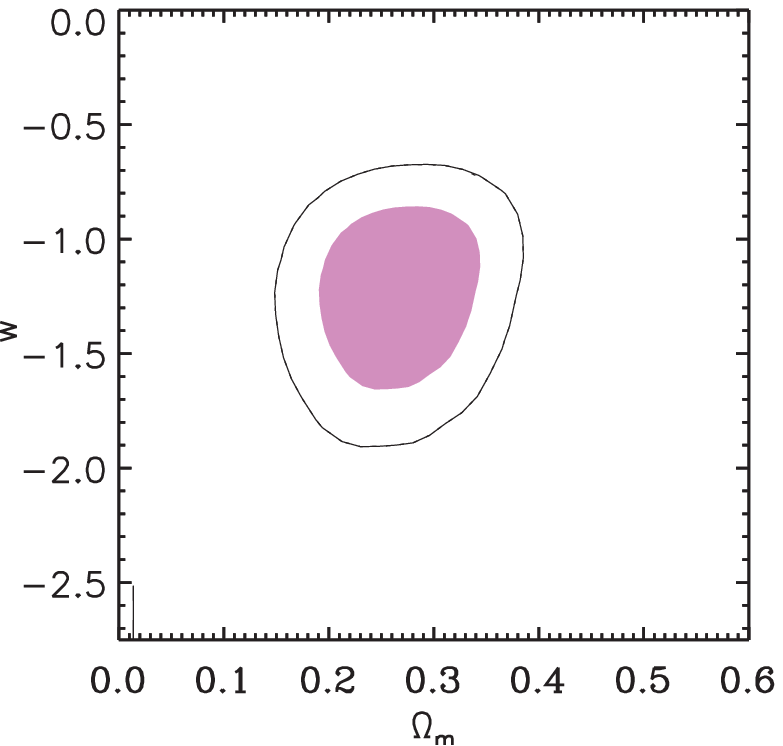,width=4.5cm}
}} 
\caption{
Panel (a). Same as Fig. 2 but for a $w$CDM model. Panel (b). Joint 68.3 and 95.4 per cent confidence constraints on $\Omega_m$ and $w$ for a constant-$w$ model using the X-ray luminosity function data and standard priors (as described in the text). 
}
\end{figure}

%{\bf 
In Fig. 3a, we set constraints for the $w$CDM model, and we plot the joint constraints on $\Omega_m$ and $\sigma_8$ from the luminosity function data using our standard priors, 
%(purple contours) are compared with those of WMAP (blue) in Fig. ??. Since the WMAP results are marginalized over $n_s$, we also show the XLF %results using our weak $n_s$ prior as dot-dashed lines, although the difference from the standard results is small. 
%
while Fig. 3b displays constraints on $\Omega_m$ and $w$ obtained independently from the XLF data.
%(purple), WMAP (blue), SNIa data (green) and cluster \fgas{} data (red). 
The marginalized results from the X-ray luminosity function data are 
$\Omega_m=0.27^{+0.07}_{-0.06}$, $\sigma_8=0.81^{+0.05}_{-0.06}$ and $w=-1.3^{+0.3}_{-0.4}$,
while in M08 they are again less tight $\Omega_m=0.24^{+0.15}_{-0.07}$, $\sigma_8=0.85^{+0.13}_{-0.20}$ and $w=-1.4^{+0.4}_{-0.7}$.
Our new XLF results are consistent 
%with each of these independent data sets, and 
with the cosmological-constant model ($w=-1$). 
%}

\begin{figure}[ht] 
\centerline{\hbox{
\psfig{file=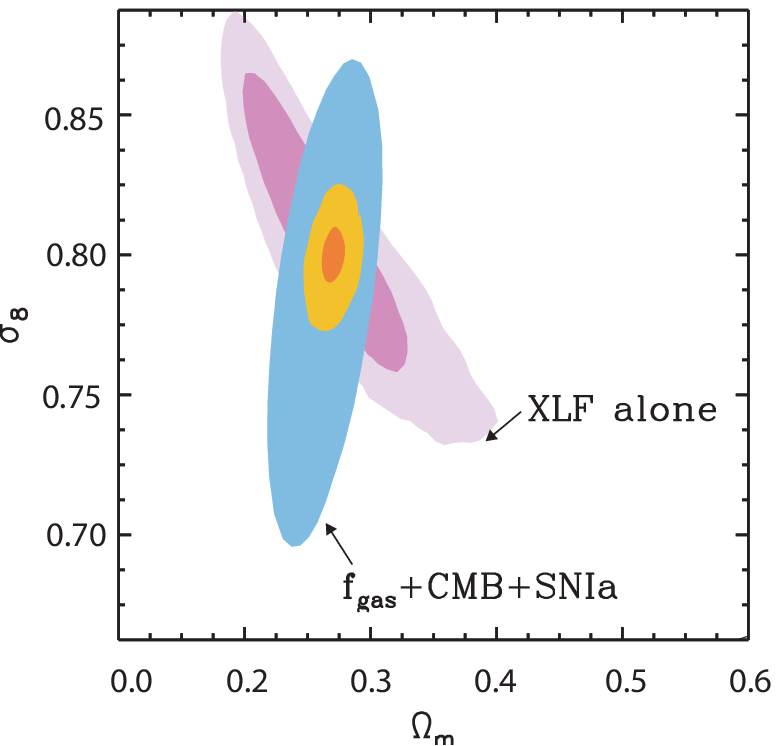,width=4.5cm}
\psfig{file=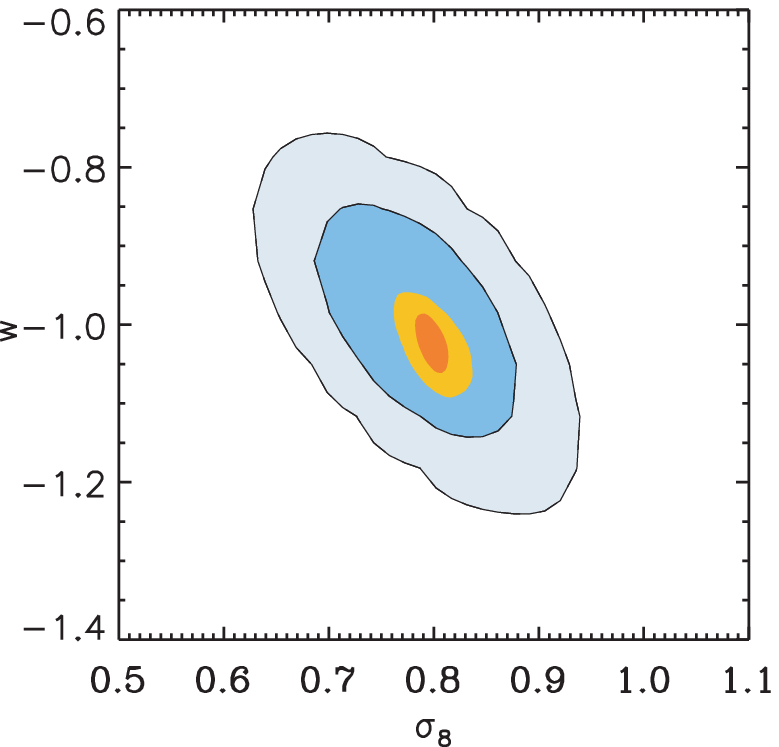,width=4.5cm}
}} 
\caption{
%%panel (c) The 68\% 
%(inner region) 
%%and 95\% 
%(outer region) 
%%credible regions of the marginalized posterior probability distribution P(w,$\Omega_m$) for a 
%%$w$CDM  model.
Joint 68.3 and 95.4 per cent confidence constraints on $\Omega_m$ and $\sigma_8$ (left panel) and $\sigma_8$ and $w$ (right panel) 
obtained from a combined $f_{gas}$+CMB+SNIa analysis (blue) and the improved constraints obtained by combining these data with the XLF (gold). 
No priors on $h$, $\Omega_b h^2$ or $n_s$ are imposed in either analysis. In the left panel, the results from the XLF alone using standard priors 
%(\tabref{tab:priors}) 
are shown (purple) in order to illustrate the degeneracy breaking.
Note that in the left panel we plotted just the inner confidence contours in the $f_{gas}$+CMB+SNIa analysis, in order to have a more readable plot.
%Also shown are independent constraints from the CMB \citep[blue;][]{Spergel07}, SNIa data \citep[green;][]{Davis07} and cluster \fgas{} data %\citep[red][]{Allen07}, and the combination of all four (gold). In the left panel, the dot-dashed lines indicate the XLF results using our weak %prior on $n_s$.}
}
\end{figure}

An improvement on the previous results can be obtained by adding to the XLF analysis the $f_{gas}$+CMB+SNIa data set. 
The $f_{gas}$+CMB+SNIa combination already provides tight constraints on $\Omega_m$, $h$, $\Omega_b h^2$ and $n_s$ (hence no priors on these parameters are used in either combined analysis), but the degeneracy between $w$ and $\sigma_8$ (right panel of Fig. 4) 
limits the precision of the dark energy results. The addition of the XLF data breaks the degeneracy in the $\Omega_m$-$\sigma_8$ plane (left panel), 
resulting in tighter constraints on $\Omega_m$, $\sigma_8$ and $w$. The degeneracy breaking power of
other combinations of data with the CMB is discussed by Spergel et al. (2007).
The resulting constraint are 
%$\Omega_m=0.269 \pm 0.015$, $\sigma_8=0.81^{\pm 0.023}$ and $w=-1.02^{\pm 0.05}$, 
$\Omega_m=0.269 \pm 0.012$, $\sigma_8=0.81 {\pm 0.021}$ and $w=-1.02 {\pm 0.04}$, 
to be compared with M08 $\Omega_m=0.269 {\pm 0.016}$, $\sigma_8=0.82 {\pm 0.03}$ and $w=-1.02 {\pm 0.06}$.
%The tight constraints for the fgas+CMB data, $H_0$ and $\Omega_b h^2$ (see Allen et al. 2007), demonstrate clearly why external
%priors on these two parameters are not required when the fgas and CMB data are combined. 
The previous constraints are in agreement within the errors with Vikhlinin et al. (2009b) constraints, namely $\Omega_m=0.255 \pm 0.043$, $\sigma_8=0.786 {\pm 0.011}$ and $w=-0.991 {\pm 0.045}$.

It is important to note that the tight constraints obtained by M08 when combining XLF analysis with the $f_{gas}$+CMB+SNIa data set are 
primarily due to the tightness of the constraints obtained from $f_{gas}$+CMB+SNIa data itself and not to the precision 
of the XLF analysis of M08, as shown by comparing our results for $\Omega_m$, $\sigma_8$, $w$ obtained by the XLF analysis with those of M08. 
%The XLF alone, as shown in Fig. 2-3 gives less tight constraints. 
In our model the improvement in the mass function model and the L-M relationship gives rise to tight constraints even when using only the XLF function. 

%{\bf 
%As mentioned above, 

In order to understand why the results of our analysis are different from those of M08, we have to stress a key point.
%some points.
The M08 paper, as well as several others papers in the literature,  
%however, that work still suffered from the fact that it 
used two different data sets: REFLEX, BCS, and MACS to constrain cosmological parameters and another external data set to constrain the luminosity--mass relation (RB02 data set), which in M08 is a power-law with three free parameters,
without explicitly accounting for selection bias. 
Consequently, it was necessary to restrict that external data set (RB02) 
%\citep{Reiprich02} 
to low redshifts and high fluxes in order to minimize the effects of selection bias, making it impossible to test for departures from self-similar evolution in the scaling relation.
%In other terms, 
In order to have a ``self-consistent'' analysis, it is necessary  that a single likelihood function be applied
%applying 
to the full data set 
%(survey + observations) 
which encompasses the entire theoretical model (cosmology + scaling relations) 
%should be derived from first principles, 
so as to ensure that the covariance among all the model parameters is fully captured and that the effects of the mass function and selection biases are properly accounted for throughout.
%More recently, 
This kind of analysis was performed for the first time by
Vikhlinin et al. (2009a,b), 
%binned their detected clusters  in redshift and mass (again with infinitesimally small bins) and 
who used the same cluster sample to constrain the scaling relations, thus obtaining tighter constraints.
%; however, their procedure still does not produce a self-consistent fit for both scaling relations and cosmology.

%
%The quoted analysis, extracting cosmological constraints and constrains to the scaling relations from a single likelihood, is necessary because 
%the scaling relations used are power laws whose parameters must be somehow fitted. 
%
In the analysis of the present paper, the L--M relation is a physically motivated relation (not a power-law with free parameters) which does not require fits to data, as in M08. 
Since we do not need 
%Again 
the double analysis of M08 and previous papers, the first to get the L--M fitting parameters from RB02 data, and the second to obtain the cosmological constraints using BCS, REFLEX and MACS, we bypass the quoted drawback in the M08 analysis. 
%is the real reason producing non tight constraints in M08
 
%NELLA LETTERA 
It is interesting to note that a month after the present paper was submitted, two papers, Mantz et al. (2009a,b), appeared in arXiv
showing that the key point that I previously stressed,
%In this section, we show that our procedure from \citetalias{Mantz08} can be 
namely generalizing M08 to allow the quoted simultaneous and self-consistent fit and using T08 mass function (instead of that in Jenkins et al. 2001)
%, using follow-up observations of flux-selected clusters to constrain the scaling relations over the full redshift range of the data, and %accounting fully for the presence of Malmquist and Eddington biases. We also show that the corresponding likelihood function can be derived from %first principles, beginning with a simple Bayesian regression model. For simplicity, we derive the likelihood for the general problem of counting %sources as a function of their properties. This general picture includes the following components:
result in cosmological constraints that are a factor 2-3 better than those in M08, based on the same flux-limited sample of clusters. 
In the present paper, we have also checked that using the same L--M relation used in M08, we reobtain the same set of constraints derived by M08\footnote{Mantz et al. (2009a,b) obtain $\Omega_m=0.27 \pm 0.02$, $\sigma_8=0.79 {\pm 0.03}$ and $w=-0.96 {\pm 0.06}$.}
.

Another point to stress concerns the use of our non-self similar L--M relation for clusters of luminosity $L> 3 \times 10^{44}$ erg/s in the 0.1--2.4 keV band. Since the clusters included in the M08 sample are high X-ray luminosity (above $3\times 10^{44}$ erg/s), one could think that 
%I would not expect 
the changes in the L--M relation of the present paper, with respect to the classical self-similar model, 
%which are primarily at gas temperatures below 3 keV (as discussed in
%earlier Del Popolo papers) 
will not produce significant changes
%to have resulted in significantly different 
in constraints on the cosmological parameters. 
Even if major differences between the L--M model of the present paper and the self--similar model are observed at 
gas temperatures below 3 ${\rm keV}$, we stress that the present L--M relation depends on the M--T and L--T relationships, 
and especially the second one (based on the Modified Punctuated Equilibrium Model (MPEM)) never behaves in a self-similar 
way as shown in Del Popolo et al. (2005) (even at gas temperatures higher than 10 keV). Moreover, as previously reported, the improvement in the constraints is 
strictly connected to the fact that we bypass the quoted drawback in the M08 analysis by means of our L--M relation not depending on parameters 
that must be fixed using external data.

In order to obtain tighter and tighter constraints one needs to try to reduce to the minimum the systematic uncertainties in the analysis. 
%The allowances for systematic uncertainties included in the analysis are relatively conservative. 
Much progress is expected over the coming years in refining the ranges of these allowances, both observationally and through improved simulations. 
A reduction in the size of the required systematic allowances will tighten the cosmological constraints.
Improved numerical simulations of large samples of massive
clusters, including a more complete treatment of star formation and feedback physics that reproduces both the observed
optical galaxy luminosity function and cluster X-ray properties, will be of major importance. 
%Progress in this area
%has been made (e.g. Bialek, Evrard & Mohr 2001, Muanwong et al. 2002, Kay S. et al. , 2004, Kravtsov, Nagai &
%Vikhlinin 2005, Ettori et al. 2004, 2006, Rasia et al. 2006;
%Nagai et al. 2007a,b), though more work remains. 
%%%%%%%%%%%%%%%%%%%%%%%%%%%%%%%%%%%%%%%%%%%%%%%%%%%%%%%%%%%%%%%%%%%%%%%%%%%%%%%%%%%%%%%%%%%%%%%%%%%%%%%%%%%In particular, this work should improve the predictions for $b(z)$.
Further deep X-ray and optical observations of nearby clusters will provide better constraints on the viscosity of the
cluster gas. Improved optical/near infrared observations of
clusters should pin down the stellar mass fraction in galaxy
clusters and its evolution.
Ground and space-based gravitational lensing studies
will provide important, independent constraints on the mass
distributions in clusters; a large program using the Subaru
telescope and Hubble Space Telescope is underway, as is similar work by other groups (e.g. Hoekstra 2007). 
%Follow-up
%observations of the SZ e?ect will also provide additional, in-
%dependent constraining power in the measurement of cosmo-
%logical parameters (the combination of direct observations of
%the SZ e?ect using radio/sub-mm data and the prediction of
%this e?ect from X-ray data provides an additional constraint
%on absolute distances to the clusters e.g. Molnar et al. 2002,
%Schmidt, Allen & Fabian 2004; Bonamente et al. 2006 and
%references therein). Moreover, the independent constraints
%provided by the SZ observations should allow a reduction
%of the priors required in future work (e.g. Rapetti & Allen
%2007).

In the near future, continuing programs of Chandra and XMM-Newton observations of known, X-ray luminous 
clusters should allow important progress to be made, both by
expanding the $f_{gas}$ sample (e.g. Chandra snapshot observations of the entire MACS sample; Ebeling et al. 2001,
2007) and through deeper observations of the current target
list. 
%The advent of new, large area SZ surveys (e.g. Ruhl
%et al. 2004) will soon provide important new target lists
%of hot, X-ray luminous high redshift clusters. 
A new, large area X-ray survey such as that proposed by the 
Spectrum-RG/eROSITA project could make a substantial contribution, 
finding hundreds of suitable systems at high redshifts.

Looking a decade ahead, the International X-ray Observatory (IXO), 
result of the merging of NASA's Constellation-X and ESA/JAXA's XEUS mission concepts, 
%Constellation-X Observatory (Con-X)
%and, later, XEUS 
will offer the possibility to carry out precise studies of dark energy using the $f_{gas}$ technique.\footnote{The large collecting area and combined
spatial/spectral resolving power of IXO
%Con-X 
should permit precise $f_{gas}$ measures. An investment of $\simeq$ 10 Ms of IXO time to measure $f_{gas}$ to 5\% (corresponding to 3.3\% accuracy in distance) in
each of the 500 hottest, most X-ray luminous, 
dynamically relaxed clusters detected in future cluster surveys, spanning the redshift range $0 < z < 2$ (typical redshift $z \simeq 0.6$),
will be sufficient to constrain cosmological parameters with a DETF figure of merit of 20–40.
%measurements with $ \simeq 5$ per cent accuracy for
%large samples ($>$ 500) of hot, massive clusters (kT $>$ 5keV)
%spanning the redshift range $0 < z < 2$ (typical redshift $z \simeq 0.6$).}. 
}

\section{Conclusions}

In the present paper, we showed how to improve the constraints on $\Omega_m$, $\sigma_8$, and the dark-energy equation-of-state parameter, $w$, 
from measurements of the X-ray luminosity function of galaxy clusters, as performed by M08. 
Improving the mass function by means of Del Popolo (2006a, b) model, which was shown to be in good agreement with T08 and using 
the L--M relationship obtained in Del Popolo (2002) and Del Popolo et al. (2005), we showed that the XLF alone can give tight constraints on
the cosmological parameters. 
Using the same methods and priors of M08, we find, for a $\Lambda$CDM universe, $\Omega_m=0.28^{+0.05}_{-0.04}$ and 
$\sigma_8=0.78^{+0.04}_{-0.05}$ and similarly in the case of a $w$CDM model, we find 
$\Omega_m=0.27^{+0.07}_{-0.06}$, $\sigma_8=0.81^{+0.05}_{-0.06}$ and $w=-1.3^{+0.3}_{-0.4}$, both tighter than M08 results. 
Combining the XLF analysis with the $f_{gas}$+CMB+SNIa data set results in the constraint 
%$\Omega_m=0.269 \pm 0.015$, $\sigma_8=0.81^{\pm 0.023}$ and $w=-1.02^{\pm 0.05}$ 
$\Omega_m=0.269 \pm 0.012$, $\sigma_8=0.81 \pm 0.021$ and $w=-1.02 \pm 0.04$, 
in agreement with the most recent determination of the quoted parameters (Allen et al. 2008; Vikhlinin et al. 2009b; Percival et al. 2009).
Our findings, consistent with $w=-1$ lends additional support to the cosmological-constant model.

\begin{acknowledgements}
The author acknowledges the financial support from the German Research Foundation (DFG) under grant NO KR 1635/16-1.
\end{acknowledgements}


\begin{thebibliography}{}
\bibitem[]{} Allen S. W., Schmidt R. W., Fabian A. C., Ebeling H., 2003, MNRAS, 342, 287
\bibitem[]{} Allen et al. 2008, MNRAS 383, Issue 3, 879
\bibitem[]{} Allen, S. W., \& Fabian, A. 1998, MNRAS, 297, L57
\bibitem[]{} Bahcall, N. A. \& Fan, X. 1998, ApJ, 504, 1
%+  (vedi astroph0208102v1)
\bibitem[]{} Bahcall,N. A., Fan, X., \& Cen, R. 1997, ApJ, 485, L53
\bibitem[]{} Bardeen J.M., Bond, J. R., Kaiser, N., Szalay, A. S. 1986, ApJ 304, 15 
\bibitem[]{} Benjamin, J. et al., 2007, MNRAS, 381, 702
\bibitem[]{} Bialek, J. J., Evrard, A. E., \& Mohr J. J. 2001, ApJ, 555, 597
\bibitem[]{} Blanchard, A., \& Bartlett, J.G. 1998,A\&A, 332, L49
\bibitem[]{} Blanchard, A., Bartlett, J. G., \& Sadat, R. 1998, preprint (astro-ph/9809182)
\bibitem[]{} B\"ohringer H. et al., 2004, A\&A, 425, 367
\bibitem[]{} Borgani, S. et al., 2001, ApJ, 561, 13
\bibitem[]{} Bower, R. G., Castander, F. J., Couch, W., Ellis, R. S., \& B\"ohringer, H. 1997, MNRAS, 291, 353
\bibitem[]{} Bryan,G. L., \& Norman,M. L. 1998,ApJ, 495, 80
\bibitem[]{} Burenin, R. A., Vikhlinin, A., Hornstrup, A., Ebeling, H., Quintana, H., \& Mescheryakov, A. 2007, ApJS, 172, 561, 
\bibitem[]{} Cavaliere, A., Menci, N., \& Tozzi, P. 1997, ApJ, 484, L21 (CMT97)
\bibitem[]{} ———. 1998, ApJ, 501, 493 (CMT98)
\bibitem[]{} ———. 1999, MNRAS, 308, 599 (CMT99)
%\bibitem[]{} Chambullu 2008....................................................................
\bibitem[]{} Cash, A., 1979, ApJ 228, 939
\bibitem[]{} Cole, S., et~al., 2005, MNRAS, 362, 505
%\bibitem[]{} Del Popolo, A., \& Gambera, M. 1998, A\&A, 337, 96
%\bibitem[]{} Del Popolo A., 2003, ApJ 599, 723
%\bibitem[]{} Del Popolo A., 2006, ApJ 637, 12
%\bibitem[]{} Del Popolo, A., A\&A 454, 17
\bibitem[]{}  David L.P., Slyz A., Jones C., Forman W., Vrtilek S.D., 1993, ApJ, 412, 479
\bibitem[]{} Del Popolo A., 2002, MNRAS 336, 81
\bibitem[]{} Del Popolo A., 2003, ApJ 599, 723
\bibitem[]{} Del Popolo A., Hiotelis S., Pe\'narrubia G., 2005, ApJ 628, 76 
\bibitem[]{} Del Popolo, A., 2006a, ApJ 637, 12
\bibitem[]{} Del Popolo, A., 2006b, AJ 131, 2367 
\bibitem[]{} Donahue,M., \& Voit, G.M. 1999,ApJ, 523, L137
\bibitem[]{} Dunkley, J. et al., 2008, arXiv:0811.4280
\bibitem[]{} Ebeling H., Edge A. C., Allen S.W., Crawford C. S., Fabian A. C., Huchra J. P., 2000, MNRAS, 318, 333
\bibitem[]{} Ebeling H., Edge A. C., Bohringer H., Allen S. W., Crawford C. S., Fabian A. C., Voges W., Huchra J. P., 1998,
\bibitem[]{} Ebeling H., Edge A. C., Henry J. P., 2001, ApJ, 553, 668
\bibitem[]{} Ebeling H., Barrett E., Donovan D., Ma C. J., Edge A. C., van Speybroeck L., 2007, ApJL, 661, L33
\bibitem[]{} Edge, A. C., \& Stewart, G. C. 1991, MNRAS, 252, 414
\bibitem[]{} Eke, V. R. et al., 1998, MNRAS, 298, 1145   
\bibitem[]{} Eke, V. R., Cole, S., \& Frenk, C. S. 1996, MNRAS, 282, 263
\bibitem[]{} Eisenstein D.~J. et~al., 2005, ApJ, 633, 560
\bibitem[]{} Finoguenov, A., Reiprich, T. H., \& B\"ohringer, H. 2001, A\&A, 368, 749
\bibitem[]{} Freedman W. L. et al., 2001, ApJ, 553, 47
\bibitem[]{} Fu, L. et al., 2008, A\&A, 479, 9
\bibitem[]{} Gao L., Navarro J. F., Cole S., Frenk C., White S. D. M., Springel V., Jenkins A., Neto A. F., 2007, MNRAS, submitted, arXiv:astro-ph/0711.0746
\bibitem[]{} Governato, F., Babul, A., Quinn, T., Tozzi, P., Baugh, C., Katz, N., \&
Lake,G. 1999,MNRAS, 307, 949
\bibitem[]{} Gregory P. C., \& Loredo T., 1992, ApJ 398, 146 
\bibitem[]{} Henry J. P., 2000, ApJ, 534, 565
\bibitem[]{} Henry J. P., 2004, ApJ, 609, 603
\bibitem[]{} Henry, J. P. \& Arnaud, K. A., 1991, ApJ, 372, 410
\bibitem[]{} Horner D.J., Mushotzky R.F., Scharf C.A., 1999, ApJ, 520, 78
\bibitem[]{} Hoekstra H., 2007, MNRAS 379, 317
\bibitem[]{} Hu W., Kravtsov A. V., 2003, ApJ, 584, 702 
\bibitem[]{} {Jenkins} A., {Frenk} C.~S., {White} S.~D.~M., {Colberg} J.~M., {Cole} S., {Evrard} A.~E., {Couchman} H.~M.~P.,  {Yoshida} N., 2001, MNRAS 321, 372
\bibitem[]{} Jeltema, T. E., Hallman, E. J., Burns, J. O., \& Motl, P. M. 2007, ApJ, in press, (ArXiv:0708.1518, 708
\bibitem[]{} Kaiser, N. 1986, MNRAS, 222, 323
\bibitem[]{} Kaiser, N. 1991, ApJ, 383, 104
\bibitem[]{} Kirkman D., Tytler D., Suzuki N., O'Meara J. M., Lubin D., 2003, ApJS, 149, 1
\bibitem[]{} Komatsu, E. et al., 2008, arXiv:0803.0547
\bibitem[]{} Kravtsov A. V., Vikhlinin A., Nagai D., 2006, ApJ, 650, 128
\bibitem[]{} Maughan B. J., 2007; ApJ 668, 772
\bibitem[]{} Mantz et al 2008, MNRAS, 387, 1179 (M08)
\bibitem[]{} Mantz et al 2009a, arXiv: 0909.3098
\bibitem[]{} Mantz et al 2009b, arXiv: 0909.3099
\bibitem[]{} Markevitch M., 1998, ApJ, 503, 77
\bibitem[]{} Morandi A., Ettori S., Moscardini L., 2007, MNRAS, 379, 518
\bibitem[]{} Muanwong, O., Thomas, P. A., Kay, S. T., Pearce, F. R., \& Couchman, H. M. P. 2001, ApJ, 552, L27
\bibitem[]{} Nagai, D., Vikhlinin, A., \& Kravtsov, A. V. 2007, ApJ, 655, 98
\bibitem[]{} Navarro, J. F., Frenk, C. S., \& White, S. D. M. 1995, MNRAS, 275, 720
\bibitem[]{} Nevalainen, J., Markevitch, M., \& Forman, W. 2000, ApJ, 532, 694
\bibitem[]{} Percival W.~J. et~al., 2007, ApJ, 657, 51
\bibitem[]{} Percival W.~J. et~al., 2009, astro-ph.CO/0907.1660
\bibitem[]{} Pierpaoli E., Scott D., White M., 2001, MNRAS, 325, 77
\bibitem[]{} Pierpaoli E., Borgani S., Scott D., White M., 2003, MNRAS, 342, 163
\bibitem[]{} Ponman, T. J., Cannon, D. B., \& Navarro, J. F. 1999, Nature, 397, 135
\bibitem[]{} Press W., \& Schechter P., 1974, ApJ 187, 425 
\bibitem[]{} Rasia, E., et al. 2006, MNRAS, 369, 2013
\bibitem[]{} Reed, D., et al. 2003, MNRAS, 346, 565 (R03)
\bibitem[]{} Reichart, D. E., Nichol, R. C., Castander, F. J., Burke, D. J., Romer, A. K., Holden, B. P., Collins, C. A., \& Ulmer, M. P. 1999b, ApJ, 518, 521
%Rosati, P., Borgani, S., & Norman, C. 2002,
\bibitem[]{} Reiprich, T. H. \& B\"ohringer, H., 2002, ApJ, 567, 716
\bibitem[]{} Rosati, P., Borgani S., and Norman, C., Ann. Rev. Astron. Astrophys 2002, 40: 539-77
\bibitem[]{} Sadat,R., Blanchard, A., \& Oukbir, J. 1998,A\&A, 329, 21
\bibitem[]{} Schuecker, P. et al., 2003, A\&A, 398, 867
\bibitem[]{} Seljak U., 2002, MNRAS, 337, 769
\bibitem[]{} Sheth R. K., \& Tormen G., 2002, MNRAS 329, 61  
\bibitem[]{} Spergel, D. N. et al., 2007, ApJS, 170, 377
\bibitem[]{} Tegmark M. et~al., 2006, Phys.\ Rev.\ D, 74, 123507
\bibitem[]{} Tinker et al. 2008, ApJ 688, 709
\bibitem[]{} Tozzi, P., \& Norman, C. 2001, ApJ, 546, 63
%arXiv:  0803.2706
\bibitem[]{} Tr\"umper J., 1993, Science, 260, 1769
\bibitem[]{} Viana, P. T. P., \& Liddle, A. R. 1999, MNRAS, 303, 535
\bibitem[]{} Voit, C.M. 2000, ApJ, 543, 113
\bibitem[]{} Vikhlinin, A., McNamara, B. R., Forman, W., Jones, C., Quintana, H., \& Hornstrup, A. 1998, ApJ 502, 558
\bibitem[]{} Vikhlinin, A. et al., 2009a, ApJ 692, 1033
\bibitem[]{} Vikhlinin, A. et al., 2009b, ApJ 692, 1060
%arXiv:0812.2720
\bibitem[]{} Voevodkin A., Vikhlinin A., 2004, ApJ, 601, 610
\bibitem[]{} Voit, C.M., \& Donahue,M. 1998,ApJ, 500, L111
\bibitem[]{} Voit, G. M., \& Bryan, G. 2001, Nature, 414, 425
\bibitem[]{} Warren, M. S., Abazajian, K., Holz, D. E., \& Teodoro, L. 2006, ApJ, 646, 881
\bibitem[]{} White S.D.M., Navarro J.F., Evrard A.E., Frenk C.S., 1993, Nature, 366, 429.
\bibitem[]{} Xu, H., Jing, G., \& Wu, X. 2001, ApJ, 553, 78
\bibitem[]{} Yahagi, H., Nagashima, M., \& Yoshii, Y. 2004, ApJ, 605, 709 (YNY04)
\bibitem[]{} Zhang, Y. Y., Finoguenov A., B\"ohringer H., Kneib J. P., Smith G. P., Czoske O., Soucail G., 2007, A\&A 467, 437
%\bibitem[]{} White, S. D. M. et al., 1993, Nature, 366, 429
%\bibitem[]{} Rosati, P., Borgani S., and Norman, C., Ann. Rev. Astron. Astrophys 2002, 40: 539-77
%\bibitem[]{} Haiman, Z. ., Mohr, J. J., \& Holder, G. P. 2001, ApJ, 553, 545
%\bibitem[]{} Vikhlinin et al '08.......................
\end{thebibliography}
\end{document}